\documentclass[pra,showpacs,twocolumn]{revtex4}
\usepackage{amsmath}
\usepackage{amssymb}
\usepackage{graphicx}% Include figure files
\usepackage{dcolumn}% Align table columns on decimal point
\usepackage{bm}% bold math
\begin{document}
\title{The Pauli principle, normal modes and superfluidity: the
emergence of collective organizational phenomena}
\author{D. K.\ Watson \\
University of Oklahoma \\
Homer L.\ Dodge Department of Physics and Astronomy \\
Norman, OK 73019}
\date{\today}

\begin{abstract}
Understanding the emergence of collective organizational 
phenomena is a major goal in many fields of physics from condensed matter
to cosmology.  Using a recently introduced manybody
perturbation formalism for fermions, we propose a mechanism for the
emergence of collective behavior, specifically superfluidity,
driven by quantum statistics and the enforcement of the Pauli principle
through the selection of normal modes.  The 
method, which is called symmetry invariant perturbation theory (SPT),
uses group theory and graphical techniques to solve the manybody Schrodinger
equation through first order exactly.  The solution at first order defines 
collective coordinates in terms of
five N-body normal modes, identified as breathing, center of mass, single
particle angular excitation, single particle radial excitation and phonon.
 A correspondence is established ``on paper'' that  enforces 
 the Pauli
principle through the assignment of specific
normal mode quantum numbers. Applied in the unitary regime, this normal mode
assignment yields  occupation only in  an extremely low frequency N-body phonon mode  
at ultralow temperatures.
 A single particle radial excitation mode at a much 
higher      frequency creates a gap that stabilizes the superfluidity at 
low temperatures.
 Coupled with the corresponding values
for the frequencies at unitarity obtained by this manybody calculation,
we obtain good agreement with experimental thermodynamic results 
including the lambda transition in the specific heat. Our results suggest
that the emergence of collective behavior in
macroscopic systems is driven by the Pauli principle and its selection of
the correct collective coordinates in the form of N-body normal modes.

\end{abstract}
%
%\pacs{03.75.Ss,31.15.xh,31.15.xp,05.30.Fk,05.70.Ce}
%
\maketitle

\section{Introduction}

Collective behavior of large systems of particles has long been an
area of great interest. Collections of particles can behave quite differently
from the complex motion of isolated particles, often acquiring
qualitatively simple forms of behavior. As pointed out by Anderson in his
treatise, ``More is Different'', ``the whole becomes not only more than but 
very different from the sum of its parts"\cite{anderson}.
 The appearance of magnetism, 
zero-viscosity in superfluids, and zero resistivity in
superconducting metals are all examples of simple behaviors that arise, 
not from
detailed microscopic forces, but from the emergence of collective
organizational phenomena. These phenomena depend on powerful and general 
principles of organization that are not well understood, but have the
potential to reveal fundamental insights into the collective behaviors of
large systems. These principles can supersede the difference between
classical and quantum physics and effect cooperative macroscopic behavior
that differs drastically from the expected individual microscopic behavior
(e.g. electron repulsion).
  Elucidating
the dynamics behind these 
principles of organization remains an important challenge in many fields of 
physics.

Studying quantum systems of identical particles
such as the ultracold gases can reveal the influence of 
  organizational principles that are due to 
quantum statistics,
which for fermions means the Pauli exclusion principle. The Pauli principle  
can provide an effective
repulsion that is dependent on particle statistics as opposed to 
interparticle interactions. In certain regimes, such as the unitarity regime
for ultracold gases, the Pauli principle can dominate the
physical interaction and control the dynamics.  When this
effect is dominant, systems exhibit collective behavior that is
universal 
as found in trapped Fermi superfluids at
unitarity. Universal behavior is also seen in 
the quark-gluon plasmas of the 
early universe, high temperature
superconductivity, and neutron stars. Knowledge of the
thermodynamics of the unitary gas thus has consequences for understanding the
equation of state in these other regimes at vastly different scales.
  Using cooling and trapping techniques, the 
atomic physics 
community has been able 
to study finite size systems of ultracold gases 
that exhibit universal properties.  
Such systems with sufficiently strong interactions behave identically on a 
scale given by the average particle separation, independent of the details of 
the short range interaction.

Understanding the
dynamics behind this large scale organization 
has been an elusive goal. The present paper seeks to address this goal by
proposing a simple, straightforward description of the connection between the
Pauli principle, the normal modes of a macroscopic system of identical 
particles and the emergence
of superfluidity.  This description is based on the symmetry invariant
perturbation method (SPT) and its exact 
first order solutions
which are the N-body normal modes. Much of the work in this approach, which is
equivalent to the work in any fully interacting manybody calculation, has 
been done ``on paper'' using group theory and graphical techniques.
  The Pauli principle is also
applied ``on paper'' by imposing restrictions on the normal mode
quantum numbers at first order in the perturbation. This 
not only enforces the Pauli principle with trivial numerical cost, but  
does so in a way that can 
directly 
reveal
large scale collective behavior through the N-body normal modes. 
The well-known phonon behavior of superfluids is seen to result
from enforcing the Pauli principle at ultracold temperatures.

To test this understanding of the dynamics, we use the SPT formalism to 
calculate thermodynamic properties 
in the 
unitary
regime including energy, entropy, and heat capacity.  We obtain
good
agreement with experiment.  In particular, the lambda transition
in the specific heat is clearly seen and agrees well with 
experimental results.  The good agreement with experiment 
supports the validity of this simple
description of the dynamics behind the emergence of collective behavior.
Driven by the enforcement of the Pauli principle, phonon normal modes and
a single particle radial excitation normal mode that becomes occupied
as the temperature increases,
create a gapped system that has the
correct thermodynamic behavior for this superfluid regime.

\section{The Partition Function}

In a recent paper\cite{partition}, we developed an approach for the determination of the 
partition
function for 
strongly-interacting identical 
fermions in ultracold regimes and applied it to a model system of 
harmonically-confined,
harmonically-interacting fermions, successfully 
calculating various thermodynamic
quantities\cite{partition}.

%These three elements -
%the manybody spectrum, the 
%statistical partitioning of the $N$ particles among the available levels,
%and the enforcement of the Pauli principle - together determine
%the behavior of mesoscopic and macroscopic Fermi systems. 

In the present study, we now apply this approach to the determination
of the partition
function and several thermodynamic quantities 
for the ultracold, strongly interacting, confined fermion 
systems in the unitary regime which have been extensively studied both
experimentally and theoretically\cite{zwierlein2,regal,jin1,jin2,jin3,jin,
kinast1,thomas,thomas1,thomas2,thomas3,thomas4,
horikoshi,grimm1,grimm2,jochim,ketterle1,nascimbene,
hu1,hu2,hu3,hu4,hu5,bulgac,burovski1,burovski2,nishida1,nishida2,nishida3,
leggett,
strinati,levin,randeria1,randeria2,randeria3,haussmann1,
haussmann2,haussmann3}. 
Strongly-interacting
systems are particularly challenging 
due to the exponential scaling of complexity which for conventional methods
scales as a function of particle number,
$N$. 
Accurate partition functions can
require millions
 of energy levels depending on, among other
things, the temperature. To date, determining the full energy spectrum
of systems with four or more particles remains a challenge\cite{daily}.

The SPT method circumvents these daunting numerical demands in several ways
\cite{FGpaper,energy,matrix_method,paperI,laingdensity,JMPpaper,test,toth,
rearrangeprl,prl,harmoniumpra,partition}.
Specifically, we are able to
rearrange the numerical work
 into analytic building blocks that allow a formulation 
that does not scale with $N$.  These analytic building blocks
have been calculated and stored previously minimizing the work needed
for new calculations.  The Pauli 
principle is applied ``on paper'' resulting in trivial numerical demands 
compared to conventional methods that explicitly enforce the antisymmetry 
of the manybody 
wave function. This method was recently 
successfully used to calculate ground state energies in the unitary regime
where results comparable in accuracy to benchmark Monte 
Carlo results for $N \le 30$ were obtained in a few seconds of computer 
time\cite{prl}.
We also performed an explicit test of our method of enforcing the Pauli principle
\cite{harmoniumpra}.

The first order SPT solution
yields a harmonic spectrum 
with five frequencies belonging to the five N-body normal modes. Similar 
to the confined ideal gas with its harmonic spectrum, the full
SPT spectrum is known for this manybody problem.
Determining the partition function for 
 systems with a completely known spectrum still presents a non-trivial problem  
due to the
 difficulty of determining the degeneracies of the manybody states and
enforcing the correct symmetry on those states
 as previous studies on confined ideal gases readily 
reveal\cite{borrmann,schmidt1,schmidt2,butts,
toms,schneider}.   

In this paper, as demonstrated in our earlier model study\cite{partition}, 
we use a conceptually different approach to the 
determination of the partition function. The Pauli
principle is applied very simply through a trivial normal mode quantum 
number assignment
for each term in the sum of states without ever obtaining the actual 
wave function. The degeneracy of each energy
level is a natural result  of doing a
straightforward partitioning of the number of energy quanta among all $N$ 
particles
into different normal mode
assignments according to the Pauli principle and collecting the statistics.
Finally, the full excitation spectrum is known through first order.

\section{Application: Unitary Fermi Gas}

 We assume an $N$-body system of fermions, $N=N_1+N_2$ with $N_1$ spin up and
$N_2$ spin down fermions such that $N_1 =N_2$, confined by a spherically symmetric
harmonic potential with 
frequency $\omega_{ho}$.
For the unitary regime, we replace the actual atom-atom
potential by an attractive square well potential of radius $R$ and
a potential depth $V_0$ 
adjusted  so the 
s-wave scattering length, $a_s$ is infinite\cite{prl}. 
%
%\begin{equation} \label{eq:vint2}
%V_{\mathtt{int}}(r_{ij})= V_0(\delta) \left[1-\tanh \left[
%\vphantom{\left(r_{ij}-\alpha-\frac{3}{D}(a-\alpha)\right)}
%\frac{1}{1-3\delta} \left(r_{ij}-3\delta R \right) \right] \right] \,,
%\end{equation}

%\noindent where $ V_o(\delta)=\frac{1}{1-3b\delta}$.
% This
%interaction becomes a square well of radius $R$ in the physical
%$D=3$\, limit.  

We apply this method to a Fermi gas in the unitary regime, 
using the full formalism, defining symmetry
coordinates from the internal displacement coordinates\cite{FGpaper,energy}
 and using the
FG method\cite{dcw}
to solve 
for the five
normal coordinates and their frequencies, $\bar{\omega}_{\mu}$.
The $N(N+1)/2$ roots, ${\bar{\omega}_{\mu}}^2$,
are highly degenerate due to the $S_N$ symmetry,
resulting in a reduction to five distinct roots that correspond
 to five irreducible representations of
$S_N$\cite{WDC} and yield five normal modes, labelled by  
${\bf 0^+, 0^-, 1^+, 1^-, 2}$\cite{FGpaper}. The ${\bf 2}$ normal modes are 
phonon, i.e. compressional modes;
  ${\bf 1^-}$ has
single particle radial behavior;  ${\bf 1^+}$ shows single particle
angular behavior; ${\bf 0^+}$  is a center-of-mass motion,
and ${\bf 0^-}$  is a symmetric breathing motion.
The energy through first order in $\delta$:
\cite{FGpaper}
\begin{equation}
\overline{E} = \overline{E}_{\infty} + \delta \Biggl[
\sum_{\renewcommand{\arraystretch}{0}
\begin{array}[t]{r@{}l@{}c@{}l@{}l} \scriptstyle \mu = \{
  & \scriptstyle \bm{0}^\pm,\hspace{0.5ex}
  & \scriptstyle \bm{1}^\pm & , %& \\
  &  \,\scriptstyle \bm{2}   \scriptstyle  \}
            \end{array}
            \renewcommand{\arraystretch}{1} }
%\hspace{-0.50em} \sum_{\mathsf{n}_{\mu}=0}^\infty
(n_{\mu}+\frac{1}{2} d_{\mu})
\bar{\omega}_{\mu} \, + \, v_o \Biggr] \,, \label{eq:E1}
\end{equation}
\noindent gives the full spectrum of excited states through the assignment
of the normal mode quantum numbers that enforce the Pauli principle.
The possible assignments are found by
relating the normal mode states 
$|n_{{\bf 0}^+},n_{{\bf 0}^-},n_{{\bf 1}^+},n_{{\bf 1}^-},n_{\bf 2}>$ to the
states of the confining potential, a spherically symmetric
three dimensional harmonic oscillator 
$(V_{\mathtt{conf}}(r_i)=\frac{1}{2}m\omega_{ho}^2{r_i}^2)$
 for which the restrictions
imposed by antisymmetry are known.  These two series of states can be related 
in the double limit  $D\to\infty$, $\omega_{ho}\to\infty$ where both
representations are valid.  Two conditions result\cite{prl,harmoniumpra}:
\begin{equation} \renewcommand{\arraystretch}{1} 
\label{eq:quanta}
2 n_{{\bf 0}^-} + 2 n_{{\bf 1}^-} =   \sum_{i=1}^N 2 \nu_i \, ,\,\,\,
2 n_{{\bf 0}^+} + 2 n_{{\bf 1}^+} + 2 n_{\bf 2} = \sum_{i=1}^N  l_i  \,
\renewcommand{\arraystretch}{1}
\end{equation}
\noindent where the radial and
 orbital angular momentum quantum numbers of the three dimensional harmonic
oscillator, $\nu_i$ and $l_i$, respectively, satisfy  $n_i = 2\nu_i + l_i$, 
with $n_i$, the energy level quanta of the ith particle defined by: 
$E=\sum_{i=1}^N\left[n_i  +\frac{3}{2}\right] \hbar\omega_{ho} =
\sum_{i=1}^N \left[(2\nu_i + l_i) +\frac{3}{2}\right] \hbar\omega_{ho}$.

\noindent These equations determine a set of possible normal mode states
$|n_{{\bf 0}^+},n_{{\bf 0}^-},n_{{\bf 1}^+},n_{{\bf 1}^-},n_{\bf 2}>$
 that are consistent with an antisymmmetric wave function
from the known set of permissible
 harmonic oscillator configurations.

% $\bar\omega_{0^+}$, 
%$\bar\omega_{0^-}$,
%$\bar\omega_{1^+}$, $\bar\omega_{1^-}$, 
%$\bar\omega_{2}$. 
The SPT energies are
obtained from Eq.~(\ref{eq:E1}) with the normal mode quanta $n_{\mu}$
determined from Eq.~(\ref{eq:quanta}) to ensure antisymmetry.
We choose quanta that correspond to
 the lowest values of the  normal mode
frequencies to yield the lowest  energy for each excited state. This results 
in occupation in $n_{\bf 2}$, the phonon mode, and in $n_{{\bf 1}^-}$, 
a single particle radial mode, which have the lowest angular and radial  frequencies 
respectively. The conditions are:
\begin{equation} \renewcommand{\arraystretch}{1} 
\label{eq:quanta3}
2 n_{{\bf 1}^-}  =  \sum_{i=1}^N 2 \nu_i ,\,\,\,\,\,\,\,
2 n_{\bf 2} = \sum_{i=1}^N  l_i \,. 
\renewcommand{\arraystretch}{1}
\end{equation}

\section{Manybody ``pairing'' in phonon normal modes: The transition from Fermi to Bose statistics}
At first order in the SPT method, the five normal modes include an
extremely low frequency, highly degenerate phonon mode. 
This phonon mode provides a  
manybody wave function resulting in cooperative, coherent behavior of the
fermions at extremely low temperatures. This model does not describe 
pairing between individual pairs of fermions, but rather sets up a 
picture of a manybody coherent wave with 
 fermions in the highly degenerate, lowest frequency mode, 
each fermion in synced motion with many other fermions 
making it impossible to 
determine which fermion is paired with which.  This type of synced manybody
motion in real space is dictated by the Pauli principle at ultralow temperatures where this
is the only mode with nonzero quanta. This ``manybody pairing'' 
 is a precursor to the two-body 
pairing (in real space)
and allows a natural transition from Fermi statistics to Bose statistics as
individual particles form pairs. Analogously, the single particle radial excitation normal
mode does not describe excitation out of an individual pair of fermions, but
rather the excitation of a single particle out of the synced motion of the
manybody
phonon mode.

\section{Thermodynamic Results}

\begin{figure}
\includegraphics[scale=0.8]{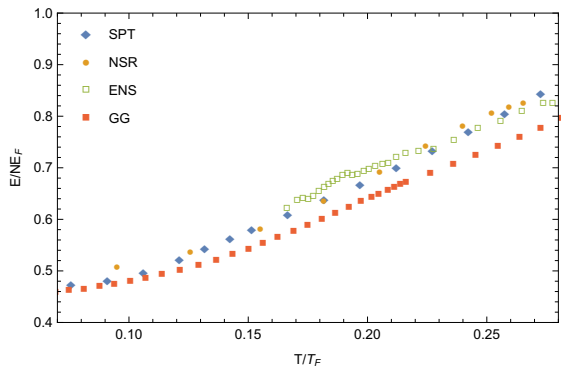}
\renewcommand{\baselinestretch}{0.8}
\caption{The universal thermodynamic function E(T). Our SPT results 
for 12 particles 
are compared to experimental: ENS\cite{nascimbene} and
theoretical results: NSR\cite{hu2,hu3} and GG\cite{hu2,hu3}. The comparison values in all the
figures were 
extracted
directly from graphs in the literature.}
\label{fig:one}
\end{figure}

\begin{figure}
\includegraphics[scale=0.8]{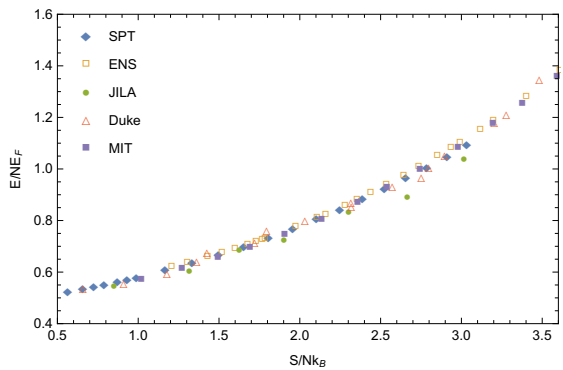}
\renewcommand{\baselinestretch}{.8}
\caption{E(S)for a trapped Fermi gas at unitarity. Our SPT results 
for 12 particles 
 are compared with experimental data: ENS\cite{nascimbene}, JILA\cite{jin1},
Duke\cite{thomas,thomas1}, MIT\cite{zwierlein2}.}

\label{fig:two}
\end{figure}

\begin{figure}
\includegraphics[scale=0.8]{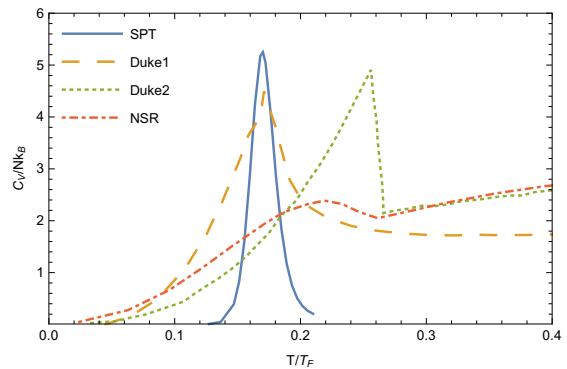}
\renewcommand{\baselinestretch}{.8}
\caption{Heat capacity vs temperature for a trapped Fermi gas at unitarity.
 Our SPT results for 36 particles
 are compared to experimental: Duke1\cite{thomas1}, Duke2\cite{kinast1}
  and theoretical results: NSR\cite{hu3}
as a function of temperature. }
\label{fig:three}
\end{figure}

We determined the following thermodynamic quantities:
energy, $E$, entropy, $S$, and heat capacity, $C_V$:

\begin{equation}
\label{eq:thermo}
\renewcommand{\arraystretch}{2.0} 
\begin{array}[b]{r@{\hspace{0.4ex}}c@{\hspace{0.2ex}}l} 
E = T^2 \frac{\partial lnZ}{\partial T},\,\,\,\,
S=S(0)+ \int_0^{T'} \, \frac{\partial E}{\partial T} \frac{1}{T} \,dT,\,\,\,\,
C_V = \frac{\partial E}{\partial T},\,\,
%\smallskip
\end{array}
\renewcommand{\arraystretch}{1}
\end{equation}

\noindent with $Z = \sum_{j=0}^\infty g_j \exp(-E_j/T)$ the
canonical partition function, $E_j$ the jth
  manybody energy, $g_j$ its degeneracy
and $T$ the temperature ($k_B=0$). 
Fig.~\ref{fig:one} shows our SPT results for the energy in units of $NE_F$,
where $E_F=(3N)^{1/3}\hbar\omega_{ho}=k_BT_F$ is the Fermi energy, compared
to experiment and theory\cite{hu2,hu3,nascimbene} as a
 function of $T/T_F$.
Our approach which does an explicit summation  included energies 
corresponding to energy quanta
up through 110.
In Fig.~\ref{fig:two} we compare our SPT results 
for  E(S), i.e. the energy vs. the entropy with   
experiment\cite{zwierlein2,nascimbene,jin1,thomas,thomas1}.
% and theory\cite{hu1,hu2,bulgac,haussmann3}. 
Finally, in  Fig.~\ref{fig:three}, we compare our SPT results for the 
heat capacity, $C_V$ in units of $Nk_B$ to previous 
results\cite{thomas1,kinast1,hu3}.

Good agreement is obtained for all three thermodynamic quantities with
 existing experimental and theoretical results.
Our calculations, which have no adjustable parameters, are the first-order
results of SPT theory. The 
calculations do show finite $N$ effects, i.e. fluctuations, as we varied $N$ 
that presumably would 
even out as $N$ increases.
As expected, below our critical temperature $T_C$, the results converge fairly
rapidly, while above $T_C$, more and more states become thermally
accessible making convergence challenging.
As $N$ is increased, the number of states needed in the partition function 
rises quite rapidly and the corresponding
degeneracies also become
quite large, straining current desktop
capabilities.  Converging the results for the particle
numbers and temperatures used in this paper was
tractable on a desktop with a few hours of time. The increase in resources
needed as $N$  and/or the temperature increases
is severe, but is not exponential. 
This is expected since we are not
obtaining the explicitly antisymmetrized wave functions for each state in a
degenerate level.  The
graph of the energy vs. temperature shows good agreement at very low 
temperatures with other theoretical
calculations. Experimental results do not yet reach these very low temperatures.

We looked
 at particle numbers between $N=10$ and $N=40$. The results in the graphs are 
for
particle number $N=12$ for Figs.~\ref{fig:one} and ~\ref{fig:two} and for $N=36$ for the heat
capacity in  Fig.~\ref{fig:three}. For E(S), we agree well with several
experimental\cite{zwierlein2,jin1,thomas,thomas1,nascimbene} and theoretical
results including NSR\cite{hu1,hu2}, Monte Carlo\cite{bulgac} and
a field theoretic approach\cite{haussmann3}.  In Fig.~\ref{fig:two}, we 
show the comparison with the experimental results.
% and include NSR results 
%to show the agreement at the lowest temperatures which are
% not yet reached by experiment. 
While the energies and entropies settled
 into good values
quickly as $N$ increased above ten; for the heat capacity,
 the lambda transition is barely
visible at $N=12$, and the critical temperature is lower $T_C\sim 0.15$,
although we
can see the convergence toward the Boltzmann value of $3Nk_B$ for a
confined gas for this low $N$.
% (See Fig.~\ref{fig:four}.)
As $N$ increases,
the lambda transition becomes quite sharp and trends to 
higher $T_C$.  
At $N=36$ shown in Fig.~\ref{fig:three}, we can converge the lambda peak, but not the higher temperature
behavior toward $3Nk_B$ 
with current desktop resources. 
We estimate the critical temperature 
to be $T_C\approx0.20T_F$. This compares well with some results in
the literature:
$(T/T_F)_C=0.19$\cite{nascimbene}, $0.20$\cite{burovski2},
 $0.21$\cite{hu2,hu5,haussmann3}, 
but is smaller than 
other reported results: $(T/T_F)_C=0.27$\cite{kinast1,hu5,bulgac}, 
0.29\cite{thomas,hu5}.

%\begin{figure}
%\includegraphics[scale=0.7]{CvN12nm100.eps}
%\renewcommand{\baselinestretch}{.8}
%\caption{Heat capacity vs temperature for a trapped Fermi gas at unitarity.
% Our SPT results for 12 particles just start to reveal a lambda peak, but
% do show convergence of this method toward the expected $3Nk_B$
% at higher temperatures.
% }
%\label{fig:four}
%\end{figure}

\section{Conclusions.} \label{sec:Conc}

Understanding the dynamics behind the emergence of collective organizational
phenomena has been a goal both experimentally and theoretically in multiple fields of physics.  In regimes exhibiting
collective behavior, simple behavior can emerge from the complexity of
the microscopic world due to the influence of organizational phenomena,         replacing microscopic uncertainty with large scale certainty.
  
From the quark-gluon plasmas of the early universe 
to ultra cold Fermi superfluids,
 the Pauli principle drives the behavior 
of mesoscopic and macroscopic Fermi systems, underpinning the emergence of
collective 
states and superseding forces at the microscopic scale. 
  Our work has
revealed that at ultralow temperatures, the Pauli principle is responsible       for selecting only a 
very low energy
phonon mode in which fermions behave in 
a cooperative manner
that minimizes the energy.
 At higher temperatures,
a single particle radial excitation normal mode becomes
occupied. This mode has a much higher frequency than the phonon mode
providing a gapped spectrum that stabilizes the superfluid behavior.

The remaining three normal modes, single excitation
angular, breathing, and center of mass, have higher
frequencies  which will become
accessible one by one depending on their magnitudes as the temperature increases.
 The exact values of these magnitudes will, of course, be influenced by the physics 
of the particular system.

These five single particle and collective modes could  also
underlie the very
successful collective and individual-particle motion description of
nuclear dynamics developed by Bohr and Mottelson\cite{bohr,bohr1} in the 
1950's which remains an important paradigm in nuclear physics.

Our picture of the physics in the unitary regime is simple, 
but it is based on a 
complex, fully-interacting solution of a manybody Hamiltonian through 
first order.
This simple physical picture emerges from the decision to use group theory to solve
the first-order harmonic equation using the FG method which
yields normal mode solutions.  For regimes with simple emergent                 behavior,
this approach provides a natural link between manybody complexity and
large scale simplicity.

The Pauli principle is known to 
have a profound effect on 
energy levels as well as quantum statistics particularly at low temperatures.
This study has now offered evidence
 that the Pauli principle has a profound effect
on the emergence of collective organizational phenomena, acting
as a powerful driving force in the emergence of collective 
organizational states of matter.  

%The stability of our hadronic universe relies fundamentally on the Pauli 
%principle and the
%Symmetrization Postulate.  For indistinguishable particles, the statistical 
%implications are known to be 
%profound. Particle statistics are found to be powerful driving
%forces in the emergence of collective organizational states of matter.  

\section{Acknowledgments}

We thank H. Hu and M. Zwierlein for providing some
detailed results. Support by the National Science Foundation through Grant 
No. PHY-1607544 is gratefully acknowledged.

\end{document}